# Enhancing Systematic Interoperability: Convergences and Mismatches between Web 3.0 and the EU Data Act

Linyi Xu
CREATe Centre, School of Law
University of Glasgow
Glasgow, UK
l.xu.3@research.gla.ac.uk

Zihao Li
CREATe Centre, School of Law
University of Glasgow
Glasgow, UK
Zihao.Li@glasgow.ac.uk

*Abstract*—Interoperability is increasingly recognised as a foundational principle for fostering innovation, competition, and user autonomy in the evolving digital ecosystem. Existing research on interoperability predominantly focuses either on technological interoperability itself or on the legal regulations concerning interoperability, with insufficient exploration of their interdisciplinary intersection. This paper compares the technological interoperability in Web 3.0 with the theoretical framework of legal interoperability established by the EU Data Act, analysing the areas of convergence and mismatch. The goal is to align technical interoperability with legal concepts of interoperability, thereby enhancing the practical implementation of systematic interoperability in the next generation of the Web. Through a comparative analysis of technological and legal interoperability, this study finds that, Firstly, Web 3.0's concept of interoperability spans data, systems, and applications, while the Data Act focuses solely on data. This narrow scope risks creating a fragmented ecosystem, where data exchange is possible, but full integration of systems and applications is hindered, leading to inefficiencies, and obstructing seamless data flow across platforms. Secondly, while Web 3.0 technically seeks to achieve interoperability through the integration of entire systems and decentralised applications, the compliance with Data Act might negatively limit such system and application interoperability through its data interoperability provisions. This paper suggests interdisciplinary recommendations to enhance the implementation and enforcement of interoperability. On one hand, the Data Act should broaden its concept of interoperability to encompass both the systems and applications layers. On the other hand, it is advisable to introduce provisions for standardised protocols through soft law mechanisms to address legal shortcomings and keep pace with technological advancements.

## I. Introduction

Compared to Web2.0, Web3.0 represents a significant technological advancement to create a more decentralized and user-centric internet [1]. Web3.0 is characterised by its emphasis on decentralisation, user data sovereignty, and smart contracts, building upon blockchain technology to enable decentralized applications (DApps) and services. This framework allows users to have control over their own data and privacy [2]. Particularly in terms of interoperability, Web3.0 facilitates communication and data sharing between applications, systems and networks, enhancing the capability to form a more unified and expansive cyber infrastructure [3]. While Web3.0 is not the only technology capable of enhancing interoperability, the decentralised architecture of Web3.0, along with its use of smart contracts, and the adoption of decentralised identity systems, undoubtedly enhances its capability to achieve interoperability.

Interoperability in Web3.0 serves as a critical foundation, facilitating seamless collaboration and data exchange across diverse platforms, systems, and applications. It denotes the efficient communication and interaction between different blockchains, decentralised applications, and protocols, ensuring cohesive operation within the Web3.0 ecosystem. In addition to technological interoperability, legal frameworks also address the concept of interoperability. Legally defined interoperability tends to focus on the theoretical aspects intended to promote and achieve interoperability through legal provisions. The concept of interoperability is firstly enshrined in the data protection legislations such as the General Data Protection Regulation (GDPR), which, through its data portability provisions, implicitly supports data sharing and flow, laying the groundwork for interoperability through a legally binding manner.

The Data Act (DA) further elaborates on the concept of interoperability, defining it as 'the ability of two or more data spaces or communication networks, systems, connected products, applications, data processing services or components to exchange and use data in order to perform their functions.' From this legal definition, it is evident that legal interoperability emphasises enhancing data flow to facilitate its use across different systems or applications. However, since DA directly addresses only data interoperability, whether it ultimately promotes interoperability among systems and applications—as defined by legal standards—requires a comparative analysis with practical implementations of interoperability.

To fully assess whether legal interoperability aligns with technological interoperability, it is crucial to compare the technical interoperability of Web3.0 with the legal requirements specified in DA.

### A. Research Questions

The main research question addressed in this study is: Can the interoperability provisions of the DA effectively facilitate comprehensive interoperability within the context of Web 3.0? This encompasses understanding the legal and technical aspects of interoperability, examining the current state and challenges, and proposing solutions to bridge the gaps.

- What are the similarities and differences between the technological interoperability of Web3.0 and the legal interoperability outlined in DA, and to what extent can Web3.0's technological interoperability be (or not) recognised by DA's requirements?

- What are the reasons and implications for the mismatch between Web3.0's technological interoperability and the legal interoperability requirements of DA, from both technical and legal perspectives, and how can technology and law be integrated to help Web3.0 comply with DA's interoperability provisions?

*B. Research Arguments and Contributions*

Although Web3.0's technological interoperability and DA's legal interoperability share some common goals, they differ significantly in terms of conceptual scope and implementation mechanisms. Web3.0 focuses on interoperability at the system and application levels, with a broader scope and more specific implementation mechanisms. In contrast, DA primarily addresses data-level interoperability and provides broader, less detailed regulatory content.

The mismatch between technological interoperability and legal interoperability arises from both technical and legal challenges. Technologically, Web3.0 must evolve within legal constraints, while legally, DA's lack of direct provisions for system and application-level interoperability, along with its limited actionable directives for promoting interoperability, contributes to this mismatch. To address these issues, DA needs to refine its concept of interoperability and enhance the interoperability of its standardisation protocols to better align with evolving Web3.0 technologies. Additionally, DA needs to enhance the feasibility of its interoperability provisions by refining specific regulations and establishing a negative list for standardization agreements, thereby defining the legal scope for practical implementation.

This study contributes to the understanding of interoperability by, first, scrutinising the interoperability provisions in DA within the context of Web3.0's technical development. Current research on interoperability is often limited to specific academic disciplines, such as competition economics, which overlooks broader interdisciplinary considerations [4]. By addressing this gap, the study moves beyond the isolated examination of technical or legal aspects. Second, it offers a comprehensive analysis of the current state of Web3.0 interoperability, highlighting areas of alignment and divergence with DA's provisions. Third, the study approaches interoperability from both technical and legal perspectives, offering interdisciplinary recommendations to improve the practical implementation of DA using Web3.0 technologies. This exploration seeks to bridge gaps between technological advancements and legal frameworks, enhancing overall interoperability.

## II. COMPARISON OF TECHNICAL AND LEGAL INTEROPERABILITY

Technical interoperability and legal interoperability are interconnected yet distinct in several ways. While both concepts focus on achieving functional integration through interoperability and emphasize effective data and information exchange, they differ in significant ways. These differences, particularly in the definition and implementation of interoperability, can lead to discrepancies where technical interoperability may conflict with or exceed the bounds of legal regulations, potentially resulting in legal non-compliance for technically robust solutions. The primary distinctions between technical and legal interoperability are evident in the definitions of interoperability and the methods used to achieve it.

*A. Conceptual Differences*

In terms of concept, the definitions of interoperability vary between the legal perspective and the technological perspective. DA aims to achieve system and application interoperability by regulating data interoperability. Its definitions and requirements regarding interoperability primarily focus on ensuring effective data exchange and utilisation across various systems and platforms. For example, Article 33 specifies essential requirements for data interoperability, including data sharing mechanisms, services, and common European data spaces. It addresses aspects such as data discovery, access, usage, and the representation of data structures. The goal is to enhance data flow and compatibility between different systems through clear provisions for data sharing and interoperability. Analysing the concept of interoperability in the DA reveals that it predominantly refers to data interoperability, emphasising the seamless exchange and use of data to achieve interoperability across systems and platforms. Beyond the DA other legal frameworks also primarily address interoperability through the lens of data interoperability. For instance, GDPR promotes data flow and interoperability through its provisions on data portability. However, its focus remains confined to the data itself. Similarly, the Public Sector Interoperability Across the Union (Interoperable Europe Act) explicitly states in Article 2 that the Act aims to enhance the sharing of data, knowledge, and information. Its emphasis is also predominantly on data interoperability.

However, technical interoperability normally encompasses a broader scope than legal interoperability. Currently, there is no consensus on the concept of technological interoperability, and it remains a topic of debate with differing perspectives. Firstly, guidelines for practitioners, particularly for computer scientists and designers, often define interoperability in terms of system interactions. For instance, the ISO/IEC/IEEE 24765 standard defines interoperability as the capability of two or more systems or components to exchange information and utilise the information that has been exchanged [5]. This implies that simple data exchange is insufficient for interoperability in systematic layer [6]. Secondly, there are perspectives that define interoperability within specific application contexts. For instance, blockchain interoperability is described as the capability of different blockchains to collaborate and operate together, rather than remaining isolated within their own environments [7]. Essentially, solutions for blockchain interoperability must allow blockchains to interact with and utilise information from other systems in a verifiable manner.

*B. Differences in Methods of Achieving Interoperability*

In terms of achieving interoperability, Web 3.0 focuses on technical methods directly applicable to systems and applications. Firstly, interoperability achieved through Web 3.0 involves the establishment and adherence to specific standards and protocols. This includes detailed and customised standards to ensure interoperability among different systems and applications, such as cross-chain interactions, DApps, standardisation and protocols, smart contract standards (e.g., ERC-20, ERC-721), and data format standards. Secondly, Web3.0 facilitates communication and asset transfers between different blockchain networks through cross-chain bridge technologies, such as Polkadot [8] and Cosmos [9], ensuring systemic integration.

In contrast, DA aims to ensure data interoperability within a legal framework by stipulating provisions for data sharing, data formats, personal data, and privacy protection. This legal framework seeks to mandate that businesses, organisations, and data service providers comply with data sharing and protection laws. It includes both general and specific provisions, with Article 33 providing fundamental regulations on data interoperability and specific obligations for data processing service providers. Additionally, the recital section of DA references the development of standardised protocols for data exchange and sharing to ensure compatibility across different systems and countries.

In summary, while Web 3.0 interoperability focuses on technical aspects and requires detailed standards for various applications, DA's interoperability provisions are more data-centric and address the needs of different stakeholders related to data. The legal framework's provisions are more fundamental and less application-specific compared to Web 3.0, leading to mismatches between the two approaches.

## III. Alignment and mismatch between Web 3.0 Technical Interoperability and Legal Interoperability

### A. Aspects of Web 3.0 Technical Interoperability Recognised by Law

Firstly, in the realm of standardisation protocols, Web 3.0's standardised protocols (e.g., ERC standards [10]) facilitate cross-platform data and asset transfers, aligning with DA's requirements for data circulation. The recitals of DA also reference the formulation of standardised protocols, highlighting their significance in the practical implementation of interoperability. Secondly, decentralised identity systems in Web 3.0 have altered data ownership mechanisms to some extent, aligning with DA's requirements for data interoperability. Web 3.0's decentralised identity systems leverage blockchain technology to give users control over their personal data, allowing them to manage and share their data autonomously. This mechanism not only enhances data security and privacy but also supports cross-platform data interoperability.

### B. Aspects of Web 3.0 Technical Interoperability That May Conflict with Legal Requirements

In addition to the aspects of technical interoperability that are already recognised by law, there are also aspects that have not yet received legal recognition. Some of these unrecognised aspects of interoperability may conflict with existing legal requirements, not only in terms of data-related laws but potentially in other areas of legislation as well. Specifically, given the variety of systems and applications within Web 3.0, some applications must adhere to both data-related regulations and those specific to their domains. As a result, efforts to achieve technical interoperability can occasionally conflict with these domain-specific regulatory requirements. For instance, applications of Web 3.0 in the financial sector must meet both interoperability standards and financial regulatory requirements. However, decentralised exchanges (DEXs) [11] in Web 3.0 do not rely on traditional financial intermediaries, which presents challenges for regulatory bodies. Traditional financial institutions are generally required to adhere strictly to financial compliance requirements, such as customer identification and KYC (Know Your Customer) regulations. The anonymity and decentralisation inherent in DEXs can make it difficult to achieve such compliance, leading to potential conflicts between the technical pursuit of interoperability and existing legal standards [12]. In addition, if a decentralised finance (DeFi) application needs to access data from users' social media accounts to verify their identity or credit score, this could conflict with privacy protection regulations and potentially infringe on users' data privacy rights.

### C. Aspects of Web 3.0 Technical Interoperability Not Yet Recognised by Law

The mismatch between technology and law can arise either from the technology itself, which may potentially conflict with legal requirements as previously mentioned, or from gaps in the existing legal framework that do not address specific technological aspects. This means that the discrepancy between the two could be due to the technology not yet being legally recognised, rather than the technology being inherently illegal. Specifically, the legal basis for Web 3.0's requirements for system and application interoperability is not addressed in DA. For example, Web 3.0 users and applications might integrate with multiple blockchain platforms, such as Ethereum. Achieving seamless data and value transfer requires addressing interoperability issues between different blockchains. However, interoperability between blockchains does not directly equate to data interoperability, and existing data regulations do not directly address this issue.

Moreover, Web 3.0 encompasses numerous DApps that may need to interact with other DApps across different blockchain platforms. While interoperability facilitates the sharing of data and functionalities, enhancing the overall ecosystem's connectivity and efficiency, the specific needs for application interoperability are not fully covered by the current data interoperability regulations. For example, consider a decentralized payment system DApp based on Ethereum that allows users to conduct transactions using cryptocurrency. At the same time, there is a decentralized NFT marketplace on the Binance Smart Chain (BSC) where users can buy and trade NFTs. If users want to use the Ethereum-based payment system to purchase NFTs on the Binance Smart Chain, it requires data and functional interoperability between the two platforms. However, current data interoperability regulations may not address the specific requirements of cross-chain payments and asset transfers, potentially leading to regulatory compliance issues during implementation.

## IV. Causes of the Mismatch Between Technical and Legal Interoperability

The mismatch between technology and law regarding interoperability may stem from issues with both the technology itself and deficiencies within the legal framework. Understanding the causes of this mismatch can help avoid outright rejection of the technology, which could hinder innovation, and provide the law with opportunities to integrate theoretical insights from practice. On the technological side, progress often outpaces legal expectations due to various innovations and approaches, which can challenge existing legal requirements. For instance, achieving interoperability may require different consensus mechanisms, data formats, and protocols. Addressing these technical discrepancies involves developing compliant and compatible protocols and tools. However, this analysis does not primarily focus on technological issues. On the legal side, the mismatch arises from the disconnect between the concept of interoperability in law and that in technology. Another contributing factor is the disparity between the legal system's need for certainty and the flexibility inherent in technological practice.

### A. Inadequacy of Legal Concepts

Legal concepts are often limited and cannot be directly equated. In the main text, DA establishes the concept of legal interoperability through three main approaches: explicitly defining interoperability, including provisions related to interoperability, and dedicating a specific chapter to interoperability. Additionally, the recitals of the DA address issues related to interoperability, providing supplementary and referential insights.

Firstly, DA provides a clear definition of interoperability through Articles 2(40) and (41). These articles define interoperability and "open interoperability specifications," underscoring the role of interoperability in enhancing data exchange efficiency by ensuring that different systems and platforms can seamlessly exchange and utilize data.

Secondly, DA addresses decentralisation provisions by examining various scenarios of data sharing from multiple perspectives and establishing regulations to promote data interoperability, as detailed in Chapters 2 through 7. For example, Chapter 6 specifically focuses on data processing services, requiring providers of cloud and edge computing services to meet interoperability standards to ensure the seamless, efficient, and swift transfer of these services. Additionally, Article 30 specifies regulations for data processing service providers regarding data portability and interoperability.

Thirdly, Chapter 8 of DA is dedicated entirely to interoperability. Articles 23 to 26 outline clear regulatory and technical parameters for effectively transitioning between data processors. The DA emphasises that all digital assets—including data, applications, virtual machines, and other forms of virtualisation technology—should be portable, including metadata generated by users. This broad definition of data demonstrates DA's comprehensive approach. Furthermore, Articles 28 and 29 establish key requirements and conditions for interoperability in data processing services, and Article 30 details essential requirements for data sharing in the context of smart contracts.

Lastly, the recitals of DA provide supplementary insights on interoperability. Although the content in recitals does not have the same legal force as the main body of the text, it plays a crucial role in refining and providing guidance on the definition and scope of interoperability. For instance, Recitals 76, 79, 80, 85, and 86 address aspects related to standardization processes. Recital 76, in particular, highlights the European Commission's view that achieving interoperability in data processing services is crucial for realizing a multi-vendor cloud environment [13]. Notably, Recital 80 suggests that standardisation could be achieved using smart contracts.

In conclusion, an analysis of the four levels of interoperability in the DA reveals that its provisions are relatively limited. Specifically, the regulations concerning system and application interoperability are broad and primarily found in the recitals, lacking the legal enforceability and practical applicability needed for effective interoperability.

### B. Lack of Feasibility in Legal Provisions for Promoting Interoperability

Firstly, legal developments may lag technological advancements, resulting in existing laws not adequately addressing the latest Web 3.0 technologies. Secondly, there are issues with the applicability of regulations. Provisions in DA may lack clarity on technical implementation details and fail to address the unique needs of decentralised technologies. For instance, standards for protocols are primarily outlined in recitals and lack legal force. Standardised protocols are crucial for promoting the free flow of data and the effective interoperability of different applications and systems. Variations in governance structures across blockchains and protocols can impact interoperability. Thus, legal assistance is necessary to manage updates and revisions to protocols and ensure consistency among all parties involved.

## V. IMPLICATIONS OF THE MISMATCH BETWEEN TECHNICAL AND LEGAL INTEROPERABILITY

### A. Consequences of the Legal Focus on Data Interoperability

When there is a significant disparity between the concepts of technological and legal interoperability, several consequences may arise. Firstly, technological interoperability encompasses a broad range of technical standards and protocols aimed at ensuring smooth interaction between different technologies and systems. In contrast, legal interoperability might focus solely on specific data exchange and processing rules. This discrepancy can result in laws failing to fully address the complex technical standards and requirements in the technology sector, thereby hindering effective regulation or promotion of practical technological applications and impacting the comprehensive implementation and cross-system compatibility of technology.

Secondly, system interoperability in practice often involves complex implementation issues and may overlook the actual integration needs of technology. Technological interoperability usually requires intricate technical solutions and standards for system integration, whereas legal interoperability might only address basic data flow regulations, data privacy, and compliance without fully considering the complexities of technical implementation. This can lead to a lack of understanding of technological implementation within the legal framework, potentially resulting in ineffective resolution of practical issues in technology deployment and operation.

### B. Consequences of the Discrepancy Between Law and Technology

When the implementation of technological interoperability is complex and detailed, while the legal promotion of interoperability steps is largely based on principles, the following impacts may occur: Firstly, technological interoperability often requires detailed technical standards and specific implementation plans to ensure effective system integration and data exchange. Legal principles, however, might lack sufficient detail to provide clear guidance for practical implementation. This can lead to situations where technological implementations fail to fully comply with legal requirements or where the law does not effectively guide and regulate the technological implementation process. Secondly, the rapid pace of technological development often outstrips the slower evolution of legal principles, which may be unable to keep up with technological advancements. This can result in the law being unable to effectively address new challenges brought by technological changes, leading to a disconnect between legal regulations and the latest technological practices.

VI. INTERDISCIPLINARY RECOMMENDATIONS FOR ENHANCING INTEROPERABILITY BETWEEN WEB 3.0 AND THE DATA ACT

As previously discussed, the mismatch between technology and law arises from both technological discrepancies and legal inadequacies. Technological non-compliance issues resulting from technological advances can be addressed through technical corrections and modifications. This means that while the development of technological interoperability needs to comply with data-related laws, it must also adhere to specific industry regulations, such as those in the financial or contractual domains. However, this paper primarily addresses the issue of legal mismatches resulting from gaps in legal provisions, which require a different approach. To bridge these gaps, the law should evolve in tandem with technological advancements by incorporating insights gained from technological practices and revising legal frameworks to better support interoperability.

### A. Establish Unified Concepts through interdisciplinary collaboration

The primary objective of interoperability is to achieve integration at both the system and application levels. System-level interoperability, compared to data interoperability, offers several distinct advantages. Firstly, it provides a stronger overall coordination, effectively advancing the ultimate goals of DA. System-level interoperability encompasses the integration and coordination of all systems and applications, including technical details such as data, protocols, and interfaces. It not only focuses on data exchange capabilities but also on the integration and optimisation of system functions, enabling different systems to operate and interact in a cohesive manner. Secondly, system-level interoperability ensures more comprehensive compatibility by guaranteeing that different systems and applications are compatible at hardware, software, and network protocol levels. This reduces barriers caused by technical architecture differences and supports a broader range of applications and system integration. Lastly, system-level interoperability can significantly enhance the efficiency of implementing interoperability. It addresses all aspects from system-level interfaces to data transmission, thereby reducing redundancy and system conflicts and optimising the overall system architecture.

However, current legal provisions, which primarily focus on facilitating data flow, fall short in achieving comprehensive system interoperability. Data interoperability typically addresses issues related to data transmission and formatting but does not tackle the challenges of efficient system-level integration and operation. Therefore, DA should expand its scope to include explicit provisions that promote the integration of systems and applications, aligning legal concepts with the broader goals of interoperability.

### B. Enhance the Practicality of Legal Provisions for Interoperability

Firstly, legal frameworks must be designed with greater practicality to avoid excessive fragmentation and ensure they do not fall behind technological developments. This can be achieved through the adoption of codes of practice that provide clear, actionable guidelines. In terms of enhancing the practicality and applicability of legal frameworks, the recent AI Act (AIA) and Digital Services Act (DSA) reflect several relevant design features. For instance, AIA categorizes artificial intelligence systems based on risk levels (e.g., high risk, low risk) and sets clear requirements and standards for each category. For high-risk AI systems, the law mandates detailed transparency, accountability, and data processing requirements (Articles 6 and 9), which indirectly support practical system interoperability. Furthermore, the AI Act includes provisions for periodic review and updates to address rapid technological advancements and emerging AI applications (Article 71). This adaptability ensures that the legal framework remains relevant and effective as technology evolves. Similarly, DSA establishes clear rules for data processing and content management for digital platforms, including mechanisms for data portability and access (Articles 13 and 14). These provisions promote interoperability between different platforms. Additionally, the DSA requires platforms to provide transparent data processing and content management policies (Article 24), supporting standardization and compatibility across various digital services. DA could benefit from the operational provisions found in AIA and DSA. By adopting similar approaches, DA could refine its interoperability clauses to enhance their feasibility and practical application.

Secondly, the DA places provisions related to the establishment of standardised protocols in the recitals, which lack legal force. While it may be challenging to immediately incorporate these provisions into the main body of the legal text, there are alternative methods to strengthen the legal framework for standardised protocols. For instance, introducing negative lists could set minimum standards for standardized protocols by clearly defining prohibited practices in their implementation. This approach allows for the integration of flexible, negotiated, or customary law elements beyond these negative provisions. Secondly, the DA addresses the establishment of standardised protocols only within the recitals, which lack legal force. While incorporating these provisions directly into the main body of the legal text may be challenging, there are alternative approaches to strengthening the legal framework for standardised protocols. One such approach is to use negative lists, which can define mandatory and universal rules for the standardisation process. For instance, the main text could specify prohibited or undesirable practices, while allowing negotiators to freely discuss and agree on other aspects not explicitly forbidden by law. This method helps close potential legal gaps that might arise from the absence of specific provisions in the legal text. For example, it could clearly prohibit practices like data monopolisation and anti-competitive behaviour, with the implication that any behaviour not explicitly prohibited is permitted under the law. By establishing prohibitive regulations for standardised protocols [14], the legal framework can better support interoperability needs in system and application domains, while also addressing potential issues related to the rigidity and lag of current laws. This would ensure that the development and implementation of standardised protocols remain legally sound and adaptable. Integrating these provisions directly into the legal text would address the current deficiencies and enhance the enforceability of standardisation requirements.

VII. CONCLUSION

DA introduces new requirements for data interoperability, establishing a novel framework for data sharing coordination among individuals, businesses, and public sectors. Its goal is to ensure that data rights holders can freely access data and stimulate the data element market through cross-sector data sharing and development. Compared to Web 2.0, Web 3.0 offers greater opportunities for interoperability, allowing different applications and platforms to interact more effectively and efficiently [15]. This results in a more integrated and cohesive digital environment. However, to achieve the goal of systematic interoperability, expanding data sharing, promoting data circulation, and facilitating transactions, the current provision in DA is not sufficient. There is a mismatch between the technological interoperability represented by Web 3.0 and the legal interoperability outlined in the DA. On one hand, the concept of legal interoperability in the DA is limited to data interoperability, neglecting the system and application interoperability required in practice. This limitation means that the law may not fully meet the technical interoperability requirements in practice. System or application-level interoperability, which excels in aspects such as comprehensive capability and compatibility, should be the primary focus. Therefore, the scope of interoperability in the DA should be broadened to include direct legal provisions for system and application-level interoperability needs. On the other hand, there is a discrepancy between the legal and technological approaches to interoperability. Currently, the law offers only principle-based provisions or references in recitals. Practical needs should be addressed by providing actionable legal guidelines. Negative lists can be used to set minimum standards for agreements and contract negotiations in practice. This approach ensures legal compliance while allowing technological development to proceed actively.

Web 3.0 is not the only technology capable of achieving interoperability, but it plays a pivotal role in its realisation. The mismatch between Web 3.0 and DA, and the process of overcoming it, provide valuable insights for both the theoretical and practical development of interoperability. By offering interdisciplinary recommendations and fostering collaborative efforts between law and technology, interoperability can be continually realised, enhancing data circulation and technological applications.